\documentclass[aps,prl,twocolumn,notitlepage,superscriptaddress]{revtex4-1}
\usepackage[colorlinks=true, citecolor=magenta, linkcolor=blue, urlcolor=blue]{hyperref}
\usepackage{graphicx}
\usepackage{amsmath}
\usepackage{amssymb}
\usepackage{amsfonts}
\usepackage{mathtools}
\usepackage{xcolor}
\usepackage{verbatim}
\usepackage{ulem}

\usepackage{tikz}
\usetikzlibrary{shapes.geometric, arrows}
\usetikzlibrary{decorations.markings}
\usetikzlibrary{decorations.pathmorphing}
\usetikzlibrary{decorations.pathreplacing}
\usetikzlibrary{shapes.symbols}
\usetikzlibrary{shapes.arrows}
\usetikzlibrary{bending}

\DeclareMathOperator{\sign}{sgn}
\DeclareMathOperator{\pf}{Pf}
\newcommand{\br}{{\bf r}}

\newcommand{\bq}{{\bf q}}

\newcommand{\bs}{{\bf s}}
\newcommand{\bS}{{\bf S}}

\allowbreak

\begin{document}

\title{Chiral superconductivity in a semiconducting wire induced by helical magnetic order}
\author{Florinda Vi\~nas Bostr\"om}
\email{florinda.bostrom@nbi.ku.dk}
\affiliation{Center for Quantum Devices, Niels Bohr Institute, University of Copenhagen, DK-2100 Copenhagen, Denmark}
\affiliation{Division of Solid State Physics and NanoLund, Lund University, Box 118, S-221 00 Lund, Sweden}
\author{Emil Vi\~nas Bostr\"om}
\email{emil.bostrom@mpsd.mpg.de}
\affiliation{Max Planck Institute for the Structure and Dynamics of Matter, Luruper Chaussee 149, 22761 Hamburg, Germany}
\date{\today}

\begin{abstract}
 Chiral superconductors are sought after for their promising but elusive Majorana zero modes. We show that a one-dimensional semiconductor in proximity to a conventional superconductor and a helical magnet can exhibit chiral superconductivity, without the need for external magnetic fields or intrinsic spin-orbit coupling. The effective proximity-induced gap and the triplet gap arising from magnon fluctuations can be made to interfere constructively. The heterostructure can be tuned into a topological regime with Majorana zero modes at its ends, over a range of chemical potentials proportional to the spin-electron coupling.
\end{abstract}

\maketitle

%%%%%%%%%%%%%%%%%%%%%%%%%%%%%%%%%%%%%%%%%%%%%%%%%%%%%%%
%%%%%%%%%%%%%%%%%%%%%%%%%%%%%%%%%%%%%%%%%%%%%%%%%%%%%%%
%%%%%%%%%%%%%%%%%%%%%%%%%%%%%%%%%%%%%%%%%%%%%%%%%%%%%%%

Superconductivity originates from the condensation of Cooper pairs, whose relative wavefunctions belong to a definite space-spin representation~\cite{Sigrist1991}. Most known superconductors arise from isotropic spin-singlet Cooper pairs, forming so-called conventional or $s$-wave superconductors. However, more exotic pairing symmetries exist, and a well known example is the cuprates, where the gap belongs to a spin-singlet representation with $d$-wave symmetry~\cite{Damascelli2003}. An even more exotic situation involves Cooper pairs of equal spins, so-called triplet pairing, whose condensation can lead to states with topological order, which are of great interest for applications in topological quantum computing~\cite{Kitaev2001,Kitaev2003,Nayak2008,Fu2008,Leijnse2012,Aasen2016}. Triplet superconductivity usually involves the breaking of chiral symmetry, associated with intrinsic time-reversal symmetry breaking and an odd parity gap function in momentum space. While intrinsic chiral superconductivity is extremely rare~\cite{Han2024}, studies suggest that it can be induced by breaking appropriate symmetries by external means~\cite{Lutchyn2010,Oreg2010,Maeland2021,Maeland2022,Maeland2023,Maeland2023b,Leraand2025,Thingstad2025}.

Previous works have attempted to engineer chiral superconductivity by combining external magnetic fields and intrinsic spin-orbit coupling~\cite{Lutchyn2010,Oreg2010,Flensberg2021}, which breaks time-reversal, spin rotation and inversion symmetry. The loss of symmetry leads to a momentum-dependent spin polarization of the electronic bands, and an effective spin-triplet pairing. Building on the same principle, platforms involving micromagnets or artificial magnetic atoms have been suggested as an alternative to semiconductors with strong intrinsic spin-orbit interaction~\cite{Braunecker2010,Kjaergaard2012,Hynes2025}. Another proposed possibility is to intercalate chiral molecules in conventional superconductors, to thereby break inversion symmetry and induce unconventional pairing in the hybrid structure. This approach was recently employed to stabilize chiral superconductivity in the layered superconductor TaS$_2$~\cite{Wan2024}. 

\begin{figure}[h]
\begin{tikzpicture}

  % Panel a
  % Wire
  \begin{scope}[shift={(1.6,4.8)},rotate around x=0,scale=0.8]

  \newcommand{\Depth}{4}
  \newcommand{\Height}{0.6}
  \newcommand{\Width}{0.5}

  \coordinate (B) at (-0.5*\Depth, 0.5*\Width, 0.5*\Height);
  \coordinate (C) at (-0.5*\Depth,-0.5*\Width, 0.5*\Height);
  \coordinate (F) at ( 0.5*\Depth, 0.5*\Width, 0.5*\Height);
  \coordinate (G) at ( 0.5*\Depth,-0.5*\Width, 0.5*\Height);
  
  \draw[violet,fill=magenta!10,opacity=0.6] (C) -- (B) -- (F) -- (G) -- cycle; % Front
  \node[text = purple] at (-0.1,-0.1) {\footnotesize Semiconductor};  
  \end{scope}
  
  % Superconductor
  \begin{scope}[shift={(1.6,5.4)},rotate around x=0,scale=0.8]
  
  \newcommand{\Depth}{4}
  \newcommand{\Height}{0.6}
  \newcommand{\Width}{0.5}

  \coordinate (B) at (-0.5*\Depth, 0.5*\Width, 0.5*\Height);
  \coordinate (C) at (-0.5*\Depth,-0.5*\Width, 0.5*\Height);
  \coordinate (F) at ( 0.5*\Depth, 0.5*\Width, 0.5*\Height);
  \coordinate (G) at ( 0.5*\Depth,-0.5*\Width, 0.5*\Height);
  
  \draw[gray!40!black,fill=gray!10,opacity=0.3] (C) -- (B) -- (F) -- (G) -- cycle; % Front
  \node[text = gray!40!black] at (-0.1,-0.1) {\footnotesize Superconductor};  
  \end{scope}
   
  \node at (1.6,3.6) {\includegraphics[width=0.45\columnwidth]{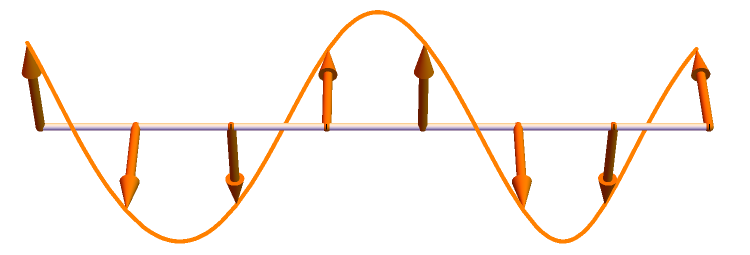}};

  \draw[<->, >=stealth, thick] (3.5,4.7) to[in=300,out=60] (3.5,5.3);
  \node at (4.15,5.20) {$\Delta_0 e^{i\theta}$};

  \draw[<->, >=stealth, thick] (3.5,3.9) to[in=300,out=60] (3.5,4.5);
  \node at (4.15,4.15) {$\Delta_{\rm mag}$};

  % Panel b
  \begin{scope}[shift={(5.45,5.25)},rotate=0,scale=0.75]
  \shade[top color=white, bottom color=white, middle color=red, shading angle=90, opacity=0.4] (0.0,-0.5) to [bend left=10] (2.5,-0.5) -- (2.5,0.5) to [bend left=10] (0.0,0.5) -- (0.0,-0.5);
  \end{scope}

  \begin{scope}[shift={(6.35,4.40)},scale=1]
  \node[single arrow, draw=blue, fill=green, minimum width = 12pt, single arrow head extend=3pt, minimum height=8mm, rotate=-90] {};
  \end{scope}

  \begin{scope}[shift={(5.45,3.43)},rotate=0,scale=0.75]
  \shade[top color=white, bottom color=white, middle color=red, shading angle=90, opacity=0.4] (0.0,-0.5) to [bend left=10] (2.5,-0.5) -- (2.5,0.5) to [bend left=10] (0.0,0.5) -- (0.0,-0.5);
  \end{scope}

  \node at (5.6,5.25) {\includegraphics[width=0.080\columnwidth]{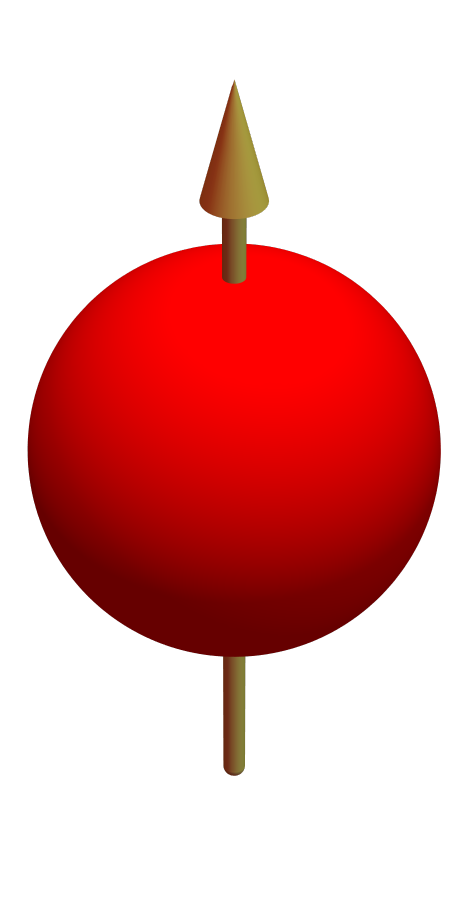}};
  \node at (7.2,5.20) {\includegraphics[width=0.082\columnwidth]{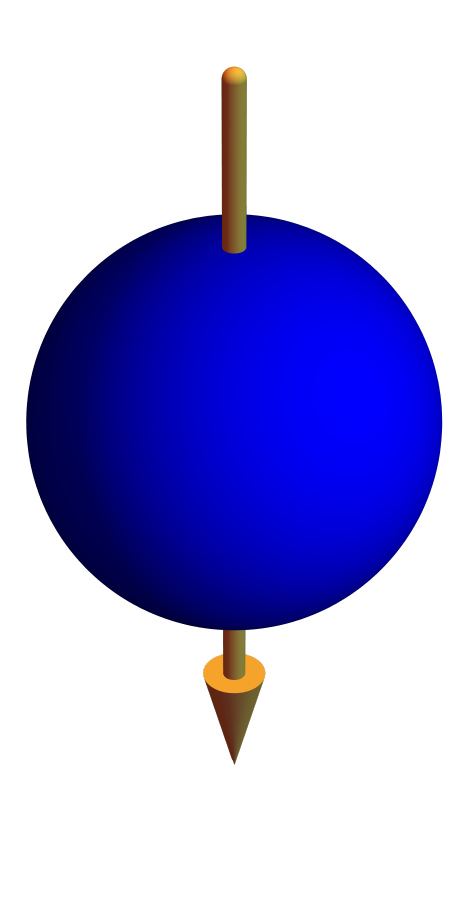}};
  \node at (5.6,3.45) {\includegraphics[width=0.080\columnwidth]{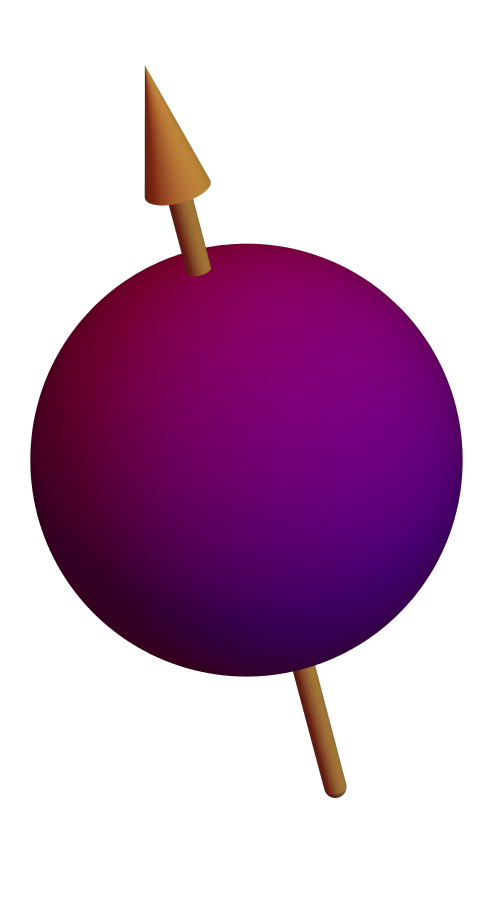}};
  \node at (7.2,3.40) {\includegraphics[width=0.082\columnwidth]{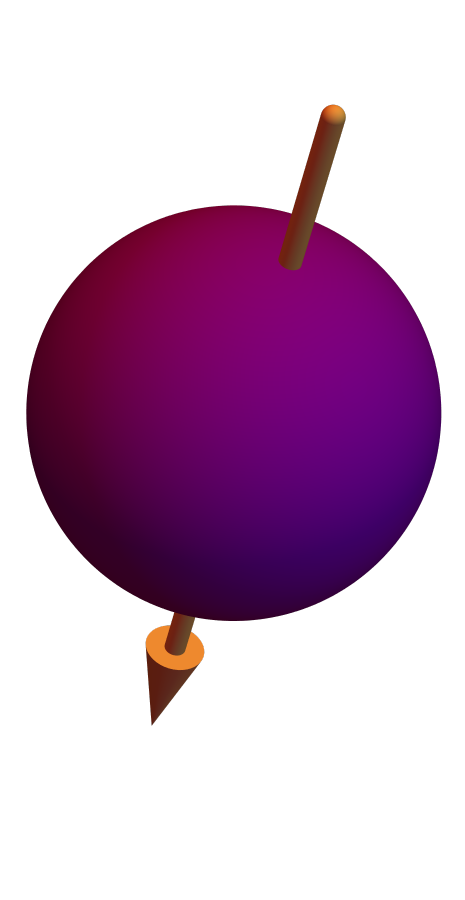}};

  %\node at (1.8,5.60) {\footnotesize Superconductor};
  %\node at (1.8,4.60) {\footnotesize Semiconductor};

  \node at (-0.45,5.60) {{\bf a}};
  \node at ( 5.00,5.65) {{\bf b}};

 \end{tikzpicture}
 \caption{({\bf a}) Three-component heterostructure consisting of a helical magnet (bottom), a semiconductor quantum  wire (middle), and a conventional superconductor (top).  The coupling between the helical magnet (orange arrows) and the semiconductor induces a magnon-mediated superconducting gap $\Delta_{\rm mag}$, while the coupling between the semiconductor and superconductor leads to a proximity-induced gap $\Delta_0 e^{i\theta}$. ({\bf b}) The non-collinear magnetic order partially transforms the conventional superconducting gap, which connects electrons with opposite spins (top), into a triplet gap connecting electrons of the same spin (bottom).}
 \label{fig:intro}
\end{figure}
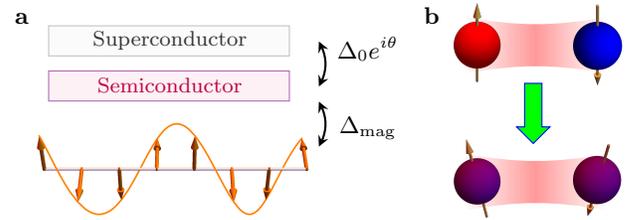

While these protocols require significant engineering or fine tuning to realize a chiral superconducting state, a promising but so far unexplored approach is to interface helical magnets with conventional superconductivity in quasi-one-dimensional (1-D) systems. Indeed, recent research into van der Waals materials has led to the discovery of short-wavelength magnetic spirals~\cite{Song2022,Gao2024,Song2025}, which can be straightforwardly interfaced with 1-D semiconductors and conventional superconductors. As demonstrated below, such structures have the advantage that chiral superconductivity can be enhanced by maximizing a single effective parameter, the spin-electron coupling $g$, thereby by-passing the need for fine tuning.

Here, we present a protocol to engineer chiral superconductivity in a three-component heterostructure, consisting of a helical magnet, a 1-D semiconductor, and a conventional superconductor (see Fig.~\ref{fig:intro}). The effective model governing this heterostructure permits an exact analytical solution, and predicts an upper limit of $\Delta/\Delta_0 = 0.5$ for the triplet gap in units of the conventional proximity-induced gap $\Delta_0$. The helical magnet breaks chiral symmetry, induces a large effective spin-orbit coupling in the semiconductor (proportional to the spiral momentum $q = 2\pi/\lambda$), and transforms the original conventional gap into an effective triplet gap in the proximitized semiconductor. Tuning the chemical potential to lie in an effective single-band regime, whose size is directly proportional to $g$, the system enters a topological superconducting state.

%%%%%%%%%%%%%%%%%%%%%%%%%%%%%%%%%%%%%%%%%%%%%%%%%%%%%%%
%%%%%%%%%%%%%%%%%%%%%%%%%%%%%%%%%%%%%%%%%%%%%%%%%%%%%%%
%%%%%%%%%%%%%%%%%%%%%%%%%%%%%%%%%%%%%%%%%%%%%%%%%%%%%%%

{\it Setup and Hamiltonian.--}
We consider a three component heterostructure consisting of a helical magnet, a 1-D semiconductor, and a conventional \textit{s}-wave superconductor, deposited on the semiconductor such that it gives rise to a proximity-induced superconducting gap in the wire (see Fig.~\ref{fig:intro}). Note that there is a large number of semiconducting materials to choose from, since the physics described does not require intrinsic spin-orbit interaction. The magnet-semiconductor system is described by the effective lattice Hamiltonian
\begin{align}\label{eq:hamiltonian}
 &H = -\mu \sum_{i\sigma} \hat{n}_{i\sigma} - t\sum_{\langle ij\rangle\sigma} \hat{c}_{i\sigma}^\dagger \hat{c}_{j\sigma} - \Delta_0 \sum_{i} (\hat{c}_{i\uparrow}^\dagger \hat{c}_{i\downarrow}^\dagger + \hat{c}_{i\uparrow} \hat{c}_{i\downarrow}) \nonumber \\
 &- J \sum_{\langle ij\rangle} \hat{\bS}_i \cdot \hat{\bS}_j - \sum_{\langle ij\rangle} {\bf D}_{ij} \cdot (\hat{\bS}_i \times \hat{\bS}_j) - g \sum_i \hat{\bs}_i \cdot \hat{\bS}_i,
\end{align}
where $\hat{n}_{i\sigma}$ is the electronic number operator, $\hat{c}_{i\sigma}$ annihilates an electron at site $i$ and of spin projection $\sigma$, and $\hat{\bS}_i$ is the operator at site $i$ for a spin of magnitude $S$. Further, the parameters $\mu$, $t$, $\Delta_0$, $J$, and ${\bf D}$ respectively determine the chemical potential, nearest-neighbor electronic hopping amplitude, the proximity-induced superconducting $s$-wave gap, the exchange interaction, and the antisymmetric Dzyaloshinskii-Moriya interaction (DMI). The electrons are assumed to live in the semiconductor, and to interact with the spins of the helical magnetic via a local spin-electron coupling of strength $g$. The electronic spin operator is defined by $\hat{\bs}_i = \sum_{\sigma\sigma'} \hat{c}_{i\sigma}^\dagger \boldsymbol\tau_{\sigma\sigma'} \hat{c}_{i\sigma'}$, with $\boldsymbol\tau$ the Pauli matrix vector.

The competition between exchange and DMI stabilizes a helical magnetic order~\cite{VinasBostrom2024}, characterized by a propagation vector $\bq$ and a vector ${\bf n}$ defining the plane of spin rotations. The magnitude of the spiral momentum is given by $q = \arctan(D/J)$, while the directions of the momentum and the rotation plane normal are determined by the vector ${\bf D}$. Here we consider a DMI such that the spiral momentum ${\bf q}$ lies along the wire axis ${\bf e}_z$, with ${\bf n}$ parallel to $\bq$. The equilibrium spin texture is a helix, and can be written as ${\bf S}_i = \cos(\bq \cdot \br_i) \, {\bf e}_x + \sin(\bq \cdot \br_i) \, {\bf e}_y$.

%%%%%%%%%%%%%%%%%%%%%%%%%%%%%%%%%%%%%%%%%%%%%%%%%%%%%%%
%%%%%%%%%%%%%%%%%%%%%%%%%%%%%%%%%%%%%%%%%%%%%%%%%%%%%%%
%%%%%%%%%%%%%%%%%%%%%%%%%%%%%%%%%%%%%%%%%%%%%%%%%%%%%%%

{\it Electronic structure with a static spin helix.--}
We first consider the effect of the static spin spiral $\bS_i$ on the electronic Hamiltonian $H_e$, consisting of the kinetic energy, the superconducting pairing term, and the spin-electron coupling. In the local coordinate frame defined by ${\bf e}_3 \parallel {\bf S}_i$, related to the global Cartesian frame by a rotation ${\bf S}_i = R_{i\alpha} {\bf S}_\alpha$, the spin-electron coupling can be written as $H_{s-e} = -g \sum_{ia\alpha} R_{a\alpha}^i \hat{s}_{ia} \hat{S}_{i\alpha}$, with $a \in \{x,y,z\}$ running over the global Cartesian axes, and $\alpha \in \{1,2,3\}$ running over the local spin axes. We focus first on the terms proportional to $\hat{S}_{i3}$, which lie along the local moments. Expanding the spin operators as $\hat{S}_{i3} = S - a_i^\dagger a_i$, and assuming that the quadratic terms can be neglected (i.e., we assume a small average magnon occupation), these terms are independent of the magnon operators $a_i$ and determine the coupling to the static spin spiral. Explicitly, the coupling becomes~\cite{VinasBostrom2024}
\begin{align}
 H_{s-e} &= -2gS \sum_{i\sigma\sigma'} \hat{c}_{i\sigma}^\dagger \begin{pmatrix} \cos\theta_i & \sin\theta_i e^{-i\phi_i} \\ \sin\theta_i e^{i\phi_l} & -\cos\theta_i \end{pmatrix} \hat{c}_{i\sigma'}
\end{align}
where $\theta_i$ and $\phi_i$ are the polar and azimuthal angles of the spin ${\bf S}_i$ in the Cartesian frame.

\begin{figure}
 \centering
 \begin{tikzpicture}
  \node at (0,0) {\includegraphics[width=1.0\columnwidth]{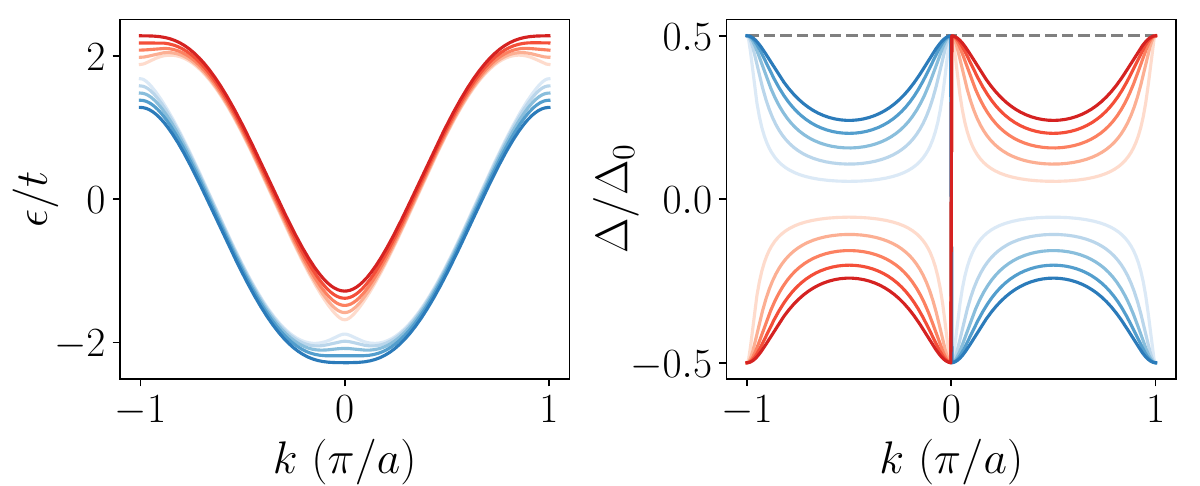}};
  \node at (-3.1,-0.7) {{\bf a}};
  \node at ( 1.8,-0.7) {{\bf b}};
 \end{tikzpicture}
 \caption{({\bf a}) Electronic band structure for electrons coupled to a spin spiral with momentum $q = 0.3\pi/a$. The band gap at $k = 0$ is given by $\Delta\epsilon = 2gS$, where $g$ is the spin-electron coupling, while the minima of the lower band are located at $k = \pm q/2$. ({\bf b}) The average singlet gap $\Delta_s = (\Delta_{du} + \Delta_{ud})/2$ (gray dashed) as well as the triplet gaps $\Delta_d(k)$ (blue) and $\Delta_u(k)$ (red). In both panels, the value of $g$ increases in steps of $0.1$ from $g = 0.1$ (lightest) to $g = 0.5$ (darkest).}
 \label{fig:gaps}
\end{figure}

For the spin spiral discussed above, propagating along the wire axis (the $z$-direction, cf. Fig.~\ref{fig:intro}) and rotating in the $xy$-plane, we have $\theta_i = \pi/2$ and $\phi_i = qr_i$. Transforming to momentum space, the electronic kinetic Hamiltonian can be cast on the form
\begin{align}\label{eq:spiral_ham}
 H_t &= \frac{1}{\sqrt{N}} \sum_k \Phi_k^\dagger \begin{pmatrix}  \epsilon_{k-\frac{q}{2}} & -gS \\ -gS & \epsilon_{k+\frac{q}{2}} \end{pmatrix} \Phi_k
\end{align}
Here $\Phi_k = (c_{k-\frac{q}{2},\uparrow} \, c_{k+\frac{q}{2},\downarrow})^T$ is a two-component spinor, and the bare electronic energies are $\epsilon_k = -2t \cos k$. With the coupling to the spiral, the energies of the electronic system are
\begin{align}
 \epsilon_{ks} &= \frac{\epsilon_{k-\frac{q}{2}}+\epsilon_{k+\frac{q}{2}}}{2} \pm \frac{1}{2}\sqrt{(\epsilon_{k+\frac{q}{2}}-\epsilon_{k-\frac{q}{2}})^2+4g^2S^2},
\end{align}
where the plus (minus) sign corresponds to pseudo-spin up (down). %$s = u$ or $d$, corresponding to $\tau_u = 1$ and $\tau_d = -1$. 
The energy bands are illustrated in Fig.~\ref{fig:gaps}a, and the operators for the corresponding eigenstates are found by a rotation of the original basis. The rotated electronic Hamiltonian (taking into account all parts of the Hamiltonian except the magnons) can be written as
\begin{align}
 H_e &= \sum_{ks} (\epsilon_{ks} -\mu) \hat{d}_{ks}^\dagger \hat{d}_{ks} - \Delta_0 \sum_{k} (\hat{c}_{k\uparrow}^\dagger \hat{c}_{-k,\downarrow}^\dagger + \hat{c}_{k\uparrow} \hat{c}_{-k,\downarrow}),
\end{align}
and what remains is to transform the basis of the pairing term. This rotation is detailed in the End Matter.

%%%%%%%%%%%%%%%%%%%%%%%%%%%%%%%%%%%%%%%%%%%%%%%%%%%%%%%
%%%%%%%%%%%%%%%%%%%%%%%%%%%%%%%%%%%%%%%%%%%%%%%%%%%%%%%
%%%%%%%%%%%%%%%%%%%%%%%%%%%%%%%%%%%%%%%%%%%%%%%%%%%%%%%

{\it Bogoliubov-de Gennes Hamiltonian.--}
Due to the spatially dependent spin rotation induced by the helical magnet, the conventional $s$-wave gap in the original basis is transformed to an effective triplet gap in the electronic band basis. The Bogoliubov-de Gennes Hamiltonian is then
\begin{align}\label{eq:bdg}
 H_e &= \frac{1}{2} \sum_k \Phi_k^\dagger \mathcal{H}_e \Phi_k, \\
 \mathcal{H}_e &= \begin{pmatrix}
                   \epsilon_{kd} - \mu & \hspace*{0.1cm}0 & \phantom{-}\Delta_d & \Delta_s \\
                   0 & \epsilon_{ku} - \mu & -\Delta_s & \Delta_u \\
                   \hspace*{0.14cm}\Delta_d & -\Delta_s & \mu - \epsilon_{kd} & \hspace*{-0.1cm}0 \\
                   \hspace*{0.14cm}\Delta_s & \phantom{-}\Delta_u & \hspace*{0.12cm}0 & \mu - \epsilon_{ku}
                  \end{pmatrix},
\end{align}
where $\Phi_k^\dagger = (\hat{d}_{kd}^\dagger,\, \hat{d}_{ku}^\dagger,\, \hat{d}_{-k,d},\, \hat{d}_{-k,u})$ and the effective gaps are related to the induced $s$-wave gap $\Delta_0$ by $\Delta_s = \Delta_0/2$,
\begin{align}
 \Delta_d(k) &= \Delta_0 \cos\frac{\theta_{k}}{2} \sin\frac{\theta_{-k}}{2},
  %\Delta_0 \sin\frac{\theta_{k}}{2} \cos\frac{\theta_{-k}}{2}
\end{align}
and $\Delta_u(k) = \Delta_d(-k)$.
The momentum dependence of the singlet and triplet gaps is shown in Fig.~\ref{fig:gaps}b, illustrating that the triplet gap is odd (as expected) while the singlet gap is even.

From the basis transformation detailed in the End Matter, we obtain the following analytical form for the triplet gaps
\begin{align}
 \Delta_d(k) &= - \frac{gS\Delta_0}{\sqrt{(\epsilon_{k+q/2}-\epsilon_{k-q/2})^2+4g^2S^2}} \sign(k) \\
 \Delta_u(k) &= \phantom{-} \frac{gS\Delta_0}{\sqrt{(\epsilon_{k+q/2}-\epsilon_{k-q/2})^2+4g^2S^2}} \sign(k). \nonumber
\end{align}
These functions are peaked at the minima of the denominator, which are located at $k = 0$ and $k = \pm \pi$ (see Fig.~\ref{fig:gaps}). The width of the peaks increase as a function of $g$, such that in the limit $g \to \infty$ the gaps are independent of momentum, with magnitude $\Delta_{u} = \Delta_{d} = \Delta_0/2$. In the opposite limit, where $\Delta\epsilon_k = |\epsilon_{k+q/2}-\epsilon_{k-q/2}| \gg gS$, the triplet gap at a given $k$ is $\Delta_{d}(k)/\Delta_0 \approx gS/\Delta\epsilon_k$.

%%%%%%%%%%%%%%%%%%%%%%%%%%%%%%%%%%%%%%%%%%%%%%%%%%%%%%%
%%%%%%%%%%%%%%%%%%%%%%%%%%%%%%%%%%%%%%%%%%%%%%%%%%%%%%%
%%%%%%%%%%%%%%%%%%%%%%%%%%%%%%%%%%%%%%%%%%%%%%%%%%%%%%%

\begin{figure}
 \centering
 \begin{tikzpicture}
  \node at (0,0) {\includegraphics[width=1.0\columnwidth]{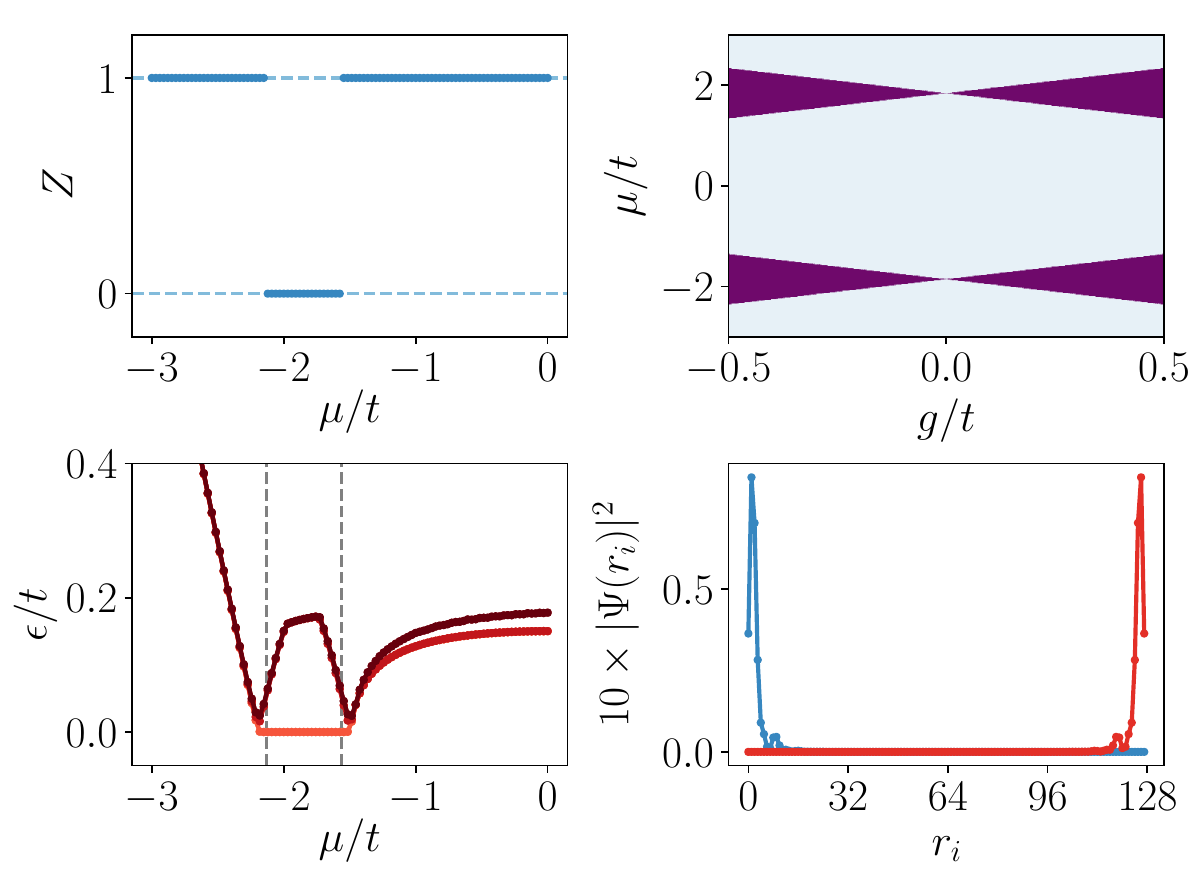}};

  \node at (-3.1, 2.2) {{\bf a}};
  \node at ( 1.4, 2.0) {{\bf b}};
  \node at (-3.1,-0.6) {{\bf c}};
  \node at ( 1.4,-0.6) {{\bf d}};
 \end{tikzpicture}
 \caption{({\bf a}) Bulk topological index as a function of chemical potential, for a spin-electron coupling $g = 0.2$. ({\bf b}) Topological phase diagram obtained from the bulk topological index $Z$. Light blue indicates the trivial phase, and dark purple indicates the topological phase. ({\bf c}) Lowest excitation energies of the finite chain as a function of chemical potential $\mu$. ({\bf d}) Wavefunctions of Majorana zero modes, obtained by exact numerical diagonalization of Eq.~\ref{eq:hamiltonian}, for a chain with $128$ sites and open boundary conditions. The chemical potential is $\mu = -1.9$, inside the bulk gap of panel ${\bf c}$. In all panels the spin spiral momentum is $q = \pi/(4a)$ and the proximity induced superconducting gap $\Delta_0 = 0.2$, with all parameters given in units of the electronic hopping $t$.}
 \label{fig:topology}
\end{figure}
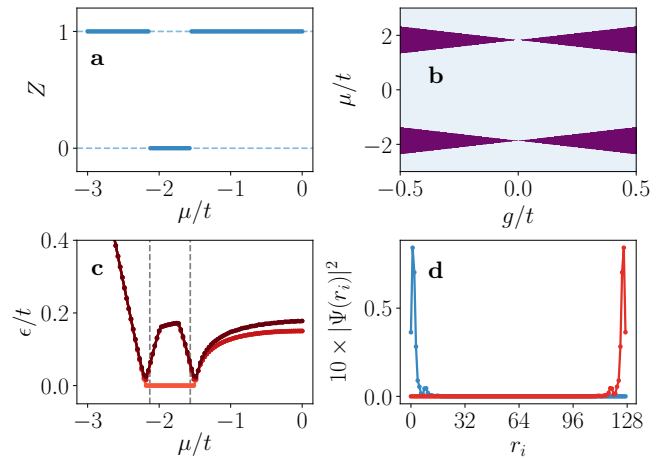

{\it Superconducting phenomenology.--}
As evidenced by these calculations and Fig.~\ref{fig:gaps}b, the singlet gap $\Delta_s$ is always greater than or equal to the triplet gaps $\Delta_d$ and $\Delta_u$. Therefore, when the chemical potential lies at an energy intersecting both electronic bands (see Fig.~\ref{fig:gaps}a), the $s$-wave gap will dominate and the wire is a conventional superconductor. In contrast, when the chemical potential lies in one of the gaps of size $2gS$, where the chemical potential intersects a single band, only the triplet pairing is efficient and the system enters a topological superconducting phase. Below, we validate this qualitative discussion by computing the bulk topological invariant, and by numerically solving Eq.~\ref{eq:hamiltonian} for a finite chain.

%%%%%%%%%%%%%%%%%%%%%%%%%%%%%%%%%%%%%%%%%%%%%%%%%%%%%%%
%%%%%%%%%%%%%%%%%%%%%%%%%%%%%%%%%%%%%%%%%%%%%%%%%%%%%%%
%%%%%%%%%%%%%%%%%%%%%%%%%%%%%%%%%%%%%%%%%%%%%%%%%%%%%%%

{\it Bulk topology.--}
A bulk topological invariant can be defined from the Pfaffian of the Hamiltonian $H$ as $Z = \sign (\pf [H(0)] \pf [H(\pi/a)])$~\cite{Kitaev2001}. Using that the Pfaffian of a general $4\times 4$ skew-symmetric matrix can be evaluated explicitly (see End Matter), we find for the present BdG Hamiltonian that $\pf[H(k)] = (\epsilon_{kd} - \mu)(\epsilon_{ku} - \mu) - \Delta_s^2$. The topological index is then
\begin{align}
 Z &= \sign( [(\epsilon_d(0) - \mu)(\epsilon_u(0) - \mu) - \Delta_s^2] \\
 &\hspace*{0.7cm} \times [(\epsilon_d(\pi/a) - \mu)(\epsilon_u(\pi/a) - \mu) - \Delta_s^2] ), \nonumber
\end{align}
and is displayed as a function of chemical potential $\mu = \epsilon_F$ and spin-electron coupling in Fig.~\ref{fig:topology}a and \ref{fig:topology}b.

For small doping, where the Fermi level is close to the bottom of the bands, only the first term of the topological index is relevant, and we can solve for the chemical potential corresponding to the topological transition. This gives
\begin{align}
 \mu = \frac{\epsilon_d(0) + \epsilon_u(0)}{2} \pm \frac{1}{2} \sqrt{(\epsilon_d(0) - \epsilon_u(0))^2 + 4\Delta^2}.
\end{align}
In the high-doping limit, where the Fermi level is closer to the top of the bands, we find an identical expression but with $k = 0$ replaced with $k = \pi/a$. These results agree very well with the boundaries of the region of chemical potential where zero modes appear for a finite length wire, as shown by Figs.~\ref{fig:topology}a and \ref{fig:topology}c.

%%%%%%%%%%%%%%%%%%%%%%%%%%%%%%%%%%%%%%%%%%%%%%%%%%%%%%%
%%%%%%%%%%%%%%%%%%%%%%%%%%%%%%%%%%%%%%%%%%%%%%%%%%%%%%%
%%%%%%%%%%%%%%%%%%%%%%%%%%%%%%%%%%%%%%%%%%%%%%%%%%%%%%%

{\it Zero modes and edge states.--}
To further validate the analytical results, we numerically diagonalize the Hamiltonian in Eq.~\ref{eq:hamiltonian} for a finite chain with open boundary conditions. We use a representative spiral momentum $qa = \pi/4$, corresponding to a spiral wavelength $L = 8a$ as approximately found in NiI$_2$~\cite{Song2022}, and consider a chain with $N = 16$ spiral periods. The total system length is then $L_{\rm tot} = NL = 128a$. We express all parameters in units of the hopping, which has a typical energy $t = 0.1 - 1$ eV, and take $g = \Delta_0 = 0.2t$.

The results are shown in Fig.~\ref{fig:topology}, where in Fig.~\ref{fig:topology}c we display the lowest positive energies of the electronic system as a function of chemical potential. In a region of chemical potentials, corresponding to $\mu$ inside the single band regime (see Fig.~\ref{fig:gaps}a), two states appear with energies pinned to zero. The wavefunctions of these states are highly localized to the edges of the chain (see Fig.~\ref{fig:topology}d), in agreement with the wavefunctions expected for topologically protected zero modes. The non-trivial topology is further corroborated by the system displaying a non-trivial bulk topological index in this range of chemical potentials. Together, the bulk and edge properties are consistent with the appearance of Majorana zero modes for $\mu$ inside the band gap.

%%%%%%%%%%%%%%%%%%%%%%%%%%%%%%%%%%%%%%%%%%%%%%%%%%%%%%%
%%%%%%%%%%%%%%%%%%%%%%%%%%%%%%%%%%%%%%%%%%%%%%%%%%%%%%%
%%%%%%%%%%%%%%%%%%%%%%%%%%%%%%%%%%%%%%%%%%%%%%%%%%%%%%%

{\it Including magnon-induced superconductivity.--}
We now include the effects of dynamical magnon fluctuations around the static spin spiral, which as discussed in our recent work~\cite{VinasBostrom2024} can give rise to superconductivity in the same way as phonons. Specifically, we consider taking the coupled semiconductor-helical magnet system, and bringing it into contact with the conventional $s$-wave superconductor. Since the magnon-induced gap is calculated in the same electronic basis as used here, it leads to a Bogoliubov-de Gennes Hamiltonian of the same form as Eq.~\ref{eq:bdg}. Neglecting the dynamical interaction of the two gaps, including the superconductivity from magnon fluctuations, therefore amounts to modifying the gap matrix of the Bogoliubov-de Gennes Hamiltonian to
\begin{align}
 \boldsymbol{\Delta} = 
  \begin{pmatrix} 
   \phantom{-}\Delta_d^{\rm mag} + e^{i\theta} \Delta_d^{\rm prox} & \Delta_s^{\rm mag} + e^{i\theta} \Delta_s^{\rm prox} \\
   -\Delta_s^{\rm mag} - e^{i\theta} \Delta_s^{\rm prox} & \Delta_u^{\rm mag} + e^{i\theta} \Delta_u^{\rm prox}
  \end{pmatrix}.
\end{align}
Here we have fixed the phase of the magnon gap to zero as a reference, and $\theta$ is the phase of the condensate in the conventional $s$-wave superconductor. 

Since the gaps arising from proximity and magnon induced superconductivity have the same form, their combined effect only depends on the phase difference $\theta$. This can be controlled by external means, e.g., by tuning the phase of the conventional superconductor using a SQUID~\cite{squid}, such that the gaps interfere constructively. Therefore, for the type of heterostructures considered here, magnon and proximity-induced superconductivity can always be made to amplify each other.

\begin{figure}
 \centering
 \begin{tikzpicture}
  \node at (0,0) {\includegraphics[width=1.0\columnwidth]{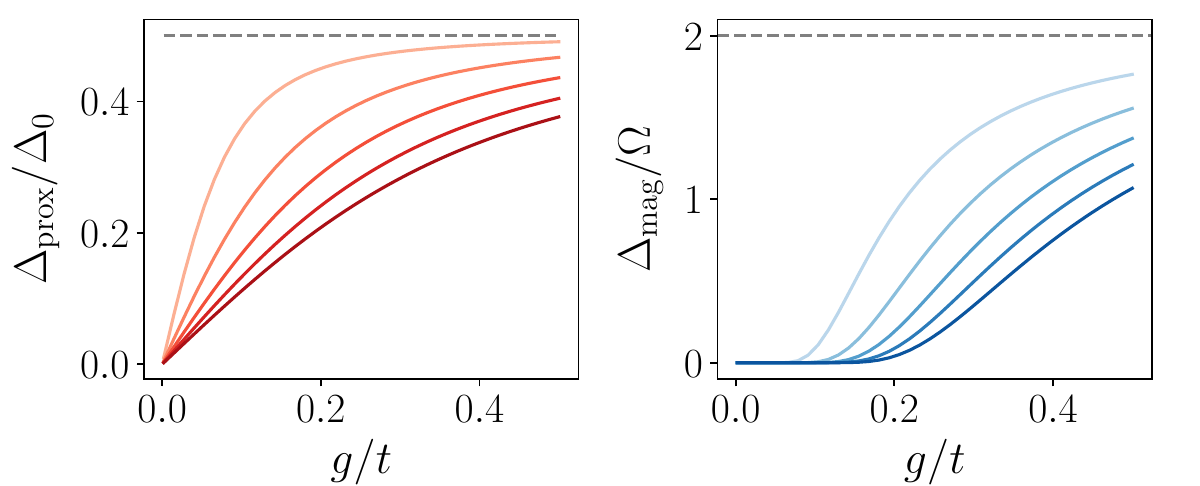}};
  \node at (-2.90, 1.2) {{\bf a}};
  \node at ( 1.25, 1.2) {{\bf b}};
 \end{tikzpicture}
 \caption{({\bf a}) Proximity induced triplet gap as a function of spin-electron coupling, for an electronic momentum $ka = 0.1\pi$. The spiral momentum increases in steps of $0.1\pi$ from $qa = 0.1\pi$ (lighest) to $qa = 0.4\pi$ (darkest). ({\bf b}) Magnon induced triplet gap as a function of spin-electron coupling, for a momentum $ka$ where the electronic dispersion is approximately linear. The magnon energy increases in steps of $0.01t$ from $\Omega/t = 0.01$ (lighest) to $\Omega/t = 0.05$ (darkest). In all panels the spin length is $S = 1$, and all parameters are given in units of the electronic hopping $t$.}
 \label{fig:gap_size}
\end{figure}

%%%%%%%%%%%%%%%%%%%%%%%%%%%%%%%%%%%%%%%%%%%%%%%%%%%%%%%
%%%%%%%%%%%%%%%%%%%%%%%%%%%%%%%%%%%%%%%%%%%%%%%%%%%%%%%
%%%%%%%%%%%%%%%%%%%%%%%%%%%%%%%%%%%%%%%%%%%%%%%%%%%%%%%

{\it Discussion and material realizations.--}
We now discuss the values of the proximity and magnon induced gaps for typical parameter values $S = 1$, $t = 1$ eV and $\Omega = 4$ meV, appropriate for the helical van der Waals magnet NiI$_2$~\cite{Gao2024}. In the effective single-band regime, the magnon gap is given by $\Delta_{\rm mag} = 2\Omega e^{-1/\lambda_{\rm mag}}$, where $\lambda_{\rm mag} \approx g^2 S/(\pi t\Omega)$ for a linear electron dispersion~\cite{VinasBostrom2024}. For a spin-electron coupling of $g = 50$ meV, the magnon-induced gap is $\Delta_{\rm mag} \approx 0.05$ meV, or equivalently $600$ mK. For the same parameters, assuming a proximity induced gap of $0.2$ meV~\cite{Microsoft2023}, we find $\Delta_{\rm prox} \approx 0.01$ meV for $q = \pi/4a$. In general, when $\Delta\epsilon_k = \epsilon_{k+q/2}-\epsilon_{k-q/2} \gg gS$, we can approximate the induced triplet gap as $\Delta/\Delta_0 \approx gS/\Delta\epsilon_{k_F}$ where $\Delta\epsilon_{k_F} \sim 2t$. The dependence of both gaps on the spin-electron coupling is illustrated in Fig.~\ref{fig:gap_size}, and shows that for small $g$ the proximity induced gap is likely dominant, since it scales linearly as opposed to exponentially with $g$.

The parameters above are appropriate when using NiI$_2$ as the magnet, and Al as the superconductor. However, we note that there is significant freedom in choosing these materials. In fact, Al is a relatively low $T_c$ superconductor, with a $T_c = 1.2$ K, while other simple materials with higher superconducting temperatures are $\Delta_{\rm Pb} = 7.2$ K, $\Delta_{\rm Nb} = 9.3$ K and $\Delta_{\rm MgB_2} = 39$ K. Therefore, swapping Al for MgB$_2$ would lead to an induced triplet gap $\Delta_{\rm prox} \approx 0.4$ meV. We also note that by utilizing moir\'e stacking the effective kinetic energy can be much reduced, such that $t$ might be as small as $t \sim 5$ meV~\cite{Cao2018}.

%%%%%%%%%%%%%%%%%%%%%%%%%%%%%%%%%%%%%%%%%%%%%%%%%%%%%%%
%%%%%%%%%%%%%%%%%%%%%%%%%%%%%%%%%%%%%%%%%%%%%%%%%%%%%%%
%%%%%%%%%%%%%%%%%%%%%%%%%%%%%%%%%%%%%%%%%%%%%%%%%%%%%%%

{\it Conclusions.--}
We have shown how to induce and enhance chiral superconductivity in a superconducting semiconductor quantum wire by placing it in the vicinity of a magnetic material with a helical spin structure. By deriving expressions for the resulting superconducting gap, we show that the effective triplet gap, as well as the range of the effective single-band regime, are both increasing functions of the same parameter, the spin-electron coupling $g$. For a phase-tunable proximate conventional superconductor, we show that the proximity-induced and magnon-induced superconducting gaps can be tuned to give additive contributions to the total effective $p$-wave gap in the semiconductor. Calculating the topological invariant, we show that the system can be stabilized within a topologically non-trivial regime, hosting Majorana zero modes at the ends in the case of open boundary conditions. 

Although we have focused on 1-D systems, we expect heterostructures involving two-dimensional (2-D) semiconductors and helical magnets to show similar unconventional and chiral properties. While the landscape of superconducting orders is significantly richer in 2-D, the proximity to a helical magnet will result in the same sort of symmetry breaking, and thereby favor chiral states. More generally, our work shows that it is possible to selectively break symmetries by proximity-coupling individual structures, and to favor a combined ground state of the heterostructure with properties inherited from its constituents. This opens a door to realize new and exotic phases in heterostructures with both twisted and direct stacking by inducing chiral symmetry breaking.

\begin{acknowledgments}
 FVB acknowledges funding from the Swedish Research Council (VR) and NanoLund, and from the European Union’s Horizon 2020 research and innovation programme under the Marie Sk{\l}odowska-Curie grant agreement No~101204715. EVB acknowledges funding from the European Union's Horizon Europe research and innovation programme under the Marie Sk{\l}odowska-Curie grant agreement No 101106809.
\end{acknowledgments}

%%%%%%%%%%%%%%%%%%%%%%%%%%%%%%%%%%%%%%%%%%%%%%%%%%%%%%%
%%%%%%%%%%%%%%%%%%%%%%%%%%%%%%%%%%%%%%%%%%%%%%%%%%%%%%%
%%%%%%%%%%%%%%%%%%%%%%%%%%%%%%%%%%%%%%%%%%%%%%%%%%%%%%%

\bibliographystyle{naturemag}

\bibliography{references}

%%%%%%%%%%%%%%%%%%%%%%%%%%%%%%%%%%%%%%%%%%%%%%%%%%%%%%%
%%%%%%%%%%%%%%%%%%%%%%%%%%%%%%%%%%%%%%%%%%%%%%%%%%%%%%%
%%%%%%%%%%%%%%%%%%%%%%%%%%%%%%%%%%%%%%%%%%%%%%%%%%%%%%%

\clearpage
%\appendix

\section*{End Matter}

\subsection{Superconducting gap in the band basis}\label{app:gap_rotation}
The corresponding eigenstates of the energies $\epsilon_{ks}$ are found by a rotation of the original basis,
\begin{align}\label{eq:transform}
  \begin{pmatrix} d_{kd} \\ d_{ku} \end{pmatrix} =
  \begin{pmatrix} \phantom{-} \cos\frac{\theta_k}{2} & \sin\frac{\theta_k}{2} \\ -\sin\frac{\theta_k}{2} & \cos\frac{\theta_k}{2} \end{pmatrix} \begin{pmatrix} c_{k-\frac{q}{2},\uparrow} \\ c_{k+\frac{q}{2},\downarrow} \end{pmatrix},
\end{align}
with the inverse transformation given by
\begin{align}\label{eq:inv_transform}
 \begin{pmatrix} c_{k\uparrow} \\ c_{k\downarrow} \end{pmatrix} =
 \begin{pmatrix} \cos\frac{\theta_{k+q/2}}{2} d_{k+\frac{q}{2},d} -\sin\frac{\theta_{k+q/2}}{2} d_{k+\frac{q}{2},u} \\ \sin\frac{\theta_{k-q/2}}{2} d_{k-\frac{q}{2},d} + \cos\frac{\theta_{k-q/2}}{2} d_{k-\frac{q}{2},u} \end{pmatrix}.
\end{align}
Here, the rotation angle $\theta_k$ is implicitly defined via the relations
\begin{subequations}
 \begin{align}
  \sin\theta_k &= \frac{2gS}{\sqrt{(\epsilon_{k+q/2}-\epsilon_{k-q/2})^2+4g^2S^2}} \\
  \cos\theta_k &= \frac{\epsilon_{k+q/2}-\epsilon_{k-q/2}}{\sqrt{(\epsilon_{k+q/2}-\epsilon_{k-q/2})^2+4g^2S^2}}.
\end{align}
\end{subequations}
Using the transformations above, we can write the electronic Hamiltonian as
\begin{align}
 H_e &= \sum_{ks} \epsilon_{ks} \hat{d}_{ks}^\dagger \hat{d}_{ks} - \Delta_0 \sum_{k} (\hat{c}_{k\uparrow}^\dagger \hat{c}_{-k,\downarrow}^\dagger + \hat{c}_{k\uparrow} \hat{c}_{-k,\downarrow}) \nonumber
\end{align}
To transform the last term, we use the rotation in Eq.~\ref{eq:inv_transform}, and make use of the fact that sum is over all $k$, such that we can relabel the summation index as $k' = k + q/2$. We then find
\begin{align}
 &\sum_{k} c_{k\uparrow} c_{-k,\downarrow} = \sum_{k} \bigg( \cos\frac{\theta_{k}}{2} \sin\frac{\theta_{-k}}{2} d_{kd} d_{-kd} \\
 &\hspace*{0.2cm}+ \cos\frac{\theta_{k}}{2} \cos\frac{\theta_{-k}}{2} d_{kd} d_{-ku} - \sin\frac{\theta_{k}}{2} \sin\frac{\theta_{-k}}{2} d_{ku} d_{-kd} \nonumber \\
 &\hspace*{0.2cm}- \sin\frac{\theta_{k}}{2} \cos\frac{\theta_{-k}}{2} d_{ku} d_{-ku} \bigg), \nonumber
\end{align}
and similarly for the complex conjugate.

%%%%%%%%%%%%%%%%%%%%%%%%%%%%%%%%%%%%%%%%%%%%%%%%%%%%%%%
%%%%%%%%%%%%%%%%%%%%%%%%%%%%%%%%%%%%%%%%%%%%%%%%%%%%%%%
%%%%%%%%%%%%%%%%%%%%%%%%%%%%%%%%%%%%%%%%%%%%%%%%%%%%%%%

\subsection{Bulk topology}\label{app:bulk_topology}
A bulk topological index can be defined from the Pfaffian of the Hamiltonian $H$, evaluated at the time-reversal invariant momenta $q = 0$ and $q = \pi/a$. It is given by $Z = \sign (\pf [H(0)] \pf [H(\pi/a)])$. To calculate the index, we note that the Pfaffian of a general $4\times 4$ skew-symmetric matrix (satisfying $A^T = -A$) can be written as $\pf(A) = af - be + dc$, provided that 
\begin{align}
 A = \begin{pmatrix} 0 &  a & b & c \\ -a &  0 &  d & e \\
                    -b & -d & 0 & f \\ -c & -e & -f & 0
     \end{pmatrix}.
\end{align}

To evaluate the topological index, we want to transform the BdG Hamiltonian into a skew-symmetric form. This is achieved by the unitary transformation
\begin{align}
 \tilde{\mathcal{H}}_e &= \frac{1}{2} \begin{pmatrix} 1 & 1 \\ i & -i \end{pmatrix}
 \begin{pmatrix} h - \mu & \Delta \\ \Delta &  \mu - h \end{pmatrix}
 \begin{pmatrix} 1 & -i \\ 1 & i \end{pmatrix} \\
 &= \begin{pmatrix} 0 & i\Delta - ih + i\mu \\ i\Delta + ih - i\mu &  0 \end{pmatrix}. \nonumber
\end{align}
This is skew-symmetric since $\Delta^T = -\Delta$, which follows from the anticommutation relations of fermionic operators and the fact that the $s$-wave ($p$-wave) gaps are even (odd) function of momentum. Writing the transformed Hamiltonian in detail, it has the form
\begin{align}
 \tilde{\mathcal{H}}_e &= \begin{pmatrix}
                   0 & 0 & i\mu - i\epsilon_{kd} & i\Delta_s \\
                   0 & 0 & -i\Delta_s & i\mu - i\epsilon_{ku} \\
                   i\epsilon_{kd} - i\mu & -i\Delta_s & 0 & 0 \\
                   i\Delta_s & i\epsilon_{ku} - i\mu & 0 & 0
                  \end{pmatrix}.
\end{align}
Here, since we are only interested in the Hamiltonian at $k = 0$ and $k = \pi/a$, we have neglected the triplet gaps that vanish at these points. The Pfaffian is now given by $\pf[H(k)] = (\epsilon_{kd} - \mu)(\epsilon_{ku} - \mu) - \Delta_s^2$, and the topological index by
\begin{align}
 Z &= \sign( [(\epsilon_d(0) - \mu)(\epsilon_u(0) - \mu) - \Delta_s^2] \\
 &\hspace*{0.7cm} \times [(\epsilon_d(\pi/a) - \mu)(\epsilon_u(\pi/a) - \mu) - \Delta_s^2] ) \nonumber
\end{align}
This is illustrated as a function of chemical potential $\mu = \epsilon_F$ in Fig.~\ref{fig:topology}c.

Noting that for small doping, where the Fermi level is close to the bottom of the bands, only the first term of the topological index is relevant, we can solve for the chemical potential corresponding to the topological transition. This gives
\begin{align}
 \mu = \frac{\epsilon_d(0) + \epsilon_u(0)}{2} \pm \frac{1}{2} \sqrt{(\epsilon_d(0) - \epsilon_u(0))^2 + 4\Delta^2}.
\end{align}
In the high-doping limit, where the Fermi level is closer to the top of the bands, we can similarly solve for the zeroes of the second term, to get
\begin{align}
 \mu = \frac{\epsilon_d(\frac{\pi}{a}) + \epsilon_u(\frac{\pi}{a})}{2} \pm \frac{1}{2} \sqrt{(\epsilon_d(\frac{\pi}{a}) - \epsilon_u(\frac{\pi}{a}))^2 + 4\Delta^2}.
\end{align}
These results agree very well with the boundaries of the region of chemical potentials where zero modes are found numerically, as shown in Fig.~\ref{fig:topology}.

%%%%%%%%%%%%%%%%%%%%%%%%%%%%%%%%%%%%%%%%%%%%%%%%%%%%%%%
%%%%%%%%%%%%%%%%%%%%%%%%%%%%%%%%%%%%%%%%%%%%%%%%%%%%%%%
%%%%%%%%%%%%%%%%%%%%%%%%%%%%%%%%%%%%%%%%%%%%%%%%%%%%%%%

\end{document}